\documentclass[twocolumn,pre,superscriptaddress,aps,showpacs,nofootinbib]{revtex4}

\usepackage[dvips]{graphicx}
\usepackage{subfigure}
\usepackage{epsfig}
\usepackage{amsmath}
\usepackage{color}

\def\be{\begin{equation}}
\def\ee{\end{equation}}
\def\bea{\begin{eqnarray}}
\def\eea{\end{eqnarray}}

\newcommand{\br}{\boldsymbol{r}}

\begin{document}
\title{Density Fluctuations in the Yukawa One Component Plasma:  \\ An accurate model for the dynamical structure factor}
\author{James P. \surname{Mithen} }\email{james.mithen@physics.ox.ac.uk}
\affiliation{Department of Physics, Clarendon Laboratory, University of Oxford, Parks Road, Oxford OX1 3PU, UK}
\author{ J\'er\^ome \surname{Daligault}}
\affiliation{Theoretical Division, Los Alamos National Laboratory, Los Alamos, NM 87545}
\author{Basil J.B. \surname{Crowley}}
\affiliation{AWE, Aldermaston, Reading RG7 4PR, UK}
\affiliation{Department of Physics, Clarendon Laboratory, University of Oxford, Parks Road, Oxford OX1 3PU, UK}
\author{Gianluca Gregori}
\affiliation{Department of Physics, Clarendon Laboratory, University of Oxford, Parks Road, Oxford OX1 3PU, UK}

\date{\today}

\begin{abstract}
Using numerical simulations, we investigate the equilibrium dynamics of a single component fluid with Yukawa interaction potential.
We show that, for a wide range of densities and temperatures, the dynamics of the system are in striking agreement with a simple model of generalized hydrodynamics.
Since the Yukawa potential can describe the ion-ion interactions in a plasma, our results have significant applicability for both analyzing and interpreting the results of x-ray scattering data from high power lasers and fourth generation light sources.
\end{abstract}


\maketitle
\section{Introduction}
Recently, using high power lasers and fourth generation x-ray sources, it has become possible to create and diagnose extreme states of matter relevant to Inertial Confinement fusion (ICF) and the cores of compact astrophysical objects in the laboratory  \cite{Glenzer,Remington,Saiz,Nagler,Kritcher}.
A particularly exciting development is that x-ray Thomson scattering experiments will soon be able to fully resolve time dependent
ion dynamics in dense plasmas \cite{Glenzer,Pelka,Gregori}.  These ion dynamics are encoded in the wavevector and frequency dependent ion-ion structure factor (or simply dynamical structure factor), $S_{ii}(k,\omega)$, which is the Fourier
transform in space and time of the density autocorrelation function.
For forthcoming experiments, an accurate model for the ion-ion structure factor is needed.

In a previous work \cite{Mithen}, we found that the conventional hydrodynamic description (Navier-Stokes equations) reproduces $S_{ii}(k,\omega)$ well for  $k < k_{max}$, where $k_{max}\lambda_s \simeq 0.43$ and 
$\lambda_s$ is the electronic screening length.  Despite the success of the conventional hydrodynamic description at these
large lengthscales (small $k$), a model that works well at higher (momentum transfer) $k$ is generally of greater applicability to the experiments.
Fortunately, a well known framework - generalized hydrodynamics - already exists for extending the results of conventional hydrodynamics to these higher $k$ values.  In this paper, we compare one of the simplest models of generalized hydrodynamics to the results of state of the art numerical simulations for $S_{ii}(k,\omega)$.
We show that the model works remarkably well for all $k$ values, 
i.e. the model describes both the conventional hydrodynamic limit at small $k$ values and the large $k$ behaviour (when the ions behave as a collection of free particles), along with the entire intermediate dynamics between these two regimes.
Our results thus show that this simple model has significant applicability for analyzing and interpreting the results of forthcoming x-ray scattering experiments using fourth generation
light sources.

This paper is structured as follows.  In Sec. \ref{simulations}, the Yukawa system - which represents interacting ions in a plasma - is introduced and details of our numerical simulations of this system are given.  In Sec. \ref{sec3}, the generalized hydrodynamics framework is summarized, along with the Gaussian approximation for the memory function that leads to a simple model for $S_{ii}(k,\omega)$.  This model is then shown to very accurately reproduce the results of our numerical simulations in Sec. \ref{results}.  
Also in this Section, we briefly discuss the applicability of our results to x-ray scattering experiments (Sec. \ref{applicability}), before offering our conclusions in Sec. \ref{conclusion}.

\section{Numerical Simulations}
\label{simulations}
We consider a plasma consisting of one species of ions of charge $Ze$ and mass $m$ at temperature $T$ and density $n$.
Because the ions are much more massive than the electrons, on the time scale of the ion dynamics of interest here, electrons instantaneously screen the ion-ion Coulomb interactions and their degrees of freedom are not treated explicitly. 
We take the Yukawa potential, 
\begin{equation}
v(r) = \frac{(Ze)^2\exp(-r/\lambda_s)}{4\pi\epsilon_0r}\,, \nonumber
\end{equation}
to represent the screened interaction between ions. 
The electronic screening length $\lambda_s$ \cite{Wunsch,Kremp,Saiz} reduces to either the Debye-Huckel law 
or the Thomas-Fermi distance in the limiting cases of classical and degenerate electron fluid respectively \cite{Glenzer}.

This single component system is known to be fully characterised by two dimensionless parameters only \cite{Donkorev}.
These are: (i) the coupling strength
\begin{equation}
\Gamma = \frac{(Ze)^2}{4\pi\epsilon_0}\frac{1}{ak_BT}\,, \nonumber
\end{equation}
where $a = (3/4\pi n)^{1/3}$ is the average inter-particle distance, and (ii) the screening parameter
\begin{equation}
\alpha = \frac{a}{\lambda_s}\,. \nonumber
\end{equation}

In our MD simulations, we compute the dynamical structure factor, $S_{ii}(k,\omega)$,
of the Yukawa system for various $\Gamma$ values ($1$,$5$,$10$,$50$,$120$,$175$) at $\alpha = 0.1,1.0 \mbox{ and } 2.0$, 
thereby spanning a range of thermodynamic conditions
\footnote{We have also performed some simulations at other $\alpha$ values; the model presented in Sec. \ref{model} works very well for these other $\alpha$ values, but here we present results for $\alpha = 0.1,1.0 \mbox{ and } 2.0$ only.}.
In our simulations, the dynamics of $N = 5000$ particles mutually interacting through the Yukawa potential are resolved using 
the Verlet algorithm in periodic boundary conditions \cite{HansenMcdonald}.
In all cases, we include the Ewald summation in our force calculation - this is essential for small $\alpha$ values - using the particle-particle-particle-mesh (PPPM) method \cite{Hockney}.  The rms error of our force calculation is $10^{-5}$.
We find that obtaining accurate MD data for $S_{ii}(k,\omega)$ requires averaging the results of a large number of
simulations to improve statistics.  This computational demand has made
a thorough study such as ours impractical before now.
For example, compared with the study of Hansen for the OCP system \cite{Hansen} - which, even after more than $30$ years remains the primary source of MD data for quantitative studies of that system \cite{Arkhipov} - we use 20 times as many particles, a smaller timestep by a factor of $2 - 10$, and simulation times $200 - 40000$ times as large.
Our timestep $\delta t = 0.01 \omega_p^{-1}$, where $\omega_p = \sqrt{(Z^2e^2n)/(\epsilon_0m)}$ is the ion plasma frequency, ensures excellent energy conservation ($\Delta E / E \approx 10^{-5}$).  
Moreover, we find that the long length  of our simulations, $25 \times 819.2\,\omega_p^{-1}$ for every $\Gamma$ and $\alpha$ value, is of paramount importance: while it is possible to capture the essential features of $S_{ii}(k,\omega)$ with simulations significantly shorter than this, producing a spectrum that is of sufficient accuracy to draw conclusions about the validity of various models requires simulations of approximately this length (we note that our data for $S_{ii}(k,\omega)$ changes negligibly by increasing the simulation time beyond $25 \times 819.2\,\omega_p^{-1}$).  In particular, these long simulation times are essential for computing accurately the decay time of collective modes at small $k$ values (i.e. the width of the ion-acoustic peak in $S_{ii}(k,\omega)$).

In a previous work \cite{Mithen}, we
presented MD results for $S_{ii}(k,\omega)$ of the Yukawa system at small $k$ values; the MD data showed that the conventional hydrodynamic description works well in describing the dynamics providing $k < k_{max}$ , where $k_{max}\lambda_{s} \simeq 0.43$.  The new MD results presented here are for a significantly larger range of $k$ values; in this paper we are interested in finding a model that reproduces the MD data for all $k$ values.
\section{Model}
\label{sec3}
\subsection{Model for $S_{ii}(k,\omega)$}
\label{model}
In the hydrodynamic regime, the wavevector and frequency dependent ion-ion structure factor can be written
\begin{equation}
\frac{S_{ii}^{H}(k,\omega)}{S_{ii}(k)} = \frac{1}{\pi}\frac{(c_sk)^2k^2\eta_l}{[\omega^2 - (c_sk)^2]^2 + [\omega k^2\eta_l]^2}\,,
\label{siihydro}
\end{equation}
where $S_{ii}(k)$ is the static ion-ion structure factor.
Equation (\ref{siihydro}) is the result obtained from the linearised Navier Stokes equation \cite{BoonYip,HansenMcdonald}.
Here $c_s$ is the (isothermal) sound speed and $\eta_l$ is the kinematic viscosity.
Equation (\ref{siihydro}) clearly has considerable similarity to the expression that underlies the model we will consider in this article
\begin{equation}
\frac{S_{ii}(k,\omega)}{S_{ii}(k)} = \frac{1}{\pi}\frac{\langle\omega_k^2\rangle k^2\phi^{'}(k,\omega)}{[\omega^2 - \langle\omega_k^2\rangle - \omega k^2 \phi^{''}(k,\omega)]^2 + [\omega k^2\phi^{'}(k,\omega)]^2}\,.
\label{siiequation}
\end{equation}
Equation (\ref{siiequation}) is a well known and exact representation of $S_{ii}(k,\omega)$ that can be formally derived from microscopic theory \cite{BalucaniZoppi}. 
The similarity to Eq. (\ref{siihydro}) is no coincidence: Eq. (\ref{siiequation}) represents a 
generalized hydrodynamics in which both equilibrium properties and transport coefficients are replaced by suitably defined
wavevector dependent quantities.
In Eq. (\ref{siiequation}), $\langle\omega_k^2\rangle = \frac{k_B T}{m}\frac{k^2}{S_{ii}(k)}$ defines a generalised isothermal sound speed $c_s(k) = \sqrt{\langle\omega_k^2\rangle/k^2} = \sqrt{\frac{k_BT}{m}\frac{1}{S_{ii}(k)}}$ that, in the hydrodynamic limit of $k \rightarrow 0$, reduces to the conventional isothermal sound speed $c_s(0) = c_s = \sqrt{\frac{k_BT}{m}\frac{\chi_T^0}{\chi_T}}$, where $\chi_T$ is the isothermal compressibility of the system and $\chi_T^0$ that of an ideal gas.
The quantities $\phi^{'}(k,\omega)$ and $\phi^{''}(k,\omega)$ are respectively the real and imaginary parts of the Laplace transform of the memory function $\phi(k,t)$: in the analogy between Eqs. (\ref{siihydro}) and (\ref{siiequation}), the memory function plays the role of a generalized viscosity. 

The model we present here amounts to using the Gaussian ansatz for the memory function,
\begin{align}
k^2\phi(k,t) &= k^2\phi(k,0)\exp(-\pi t^2/4\tau_k^2) \nonumber \\
             &= [\omega_L^2(k) - \langle\omega_k^2\rangle]\exp(-\pi t^2/4\tau_k^2)\,,
\label{Gaussianansatz}
\end{align}
where $\omega_L^2(k) = \langle\omega^4\rangle/\langle\omega^2\rangle$ is given in terms of the frequency moments of $S_{ii}(k,\omega)$
\begin{equation}
\langle\omega^n\rangle = \int_{-\infty}^{\infty}\omega^nS_{ii}(k,\omega)d\omega\,.
\end{equation}
Explicit expressions for $\langle\omega^0\rangle$, $\langle\omega^2\rangle$ and $\langle\omega^4\rangle$ are given in the Appendix. Here $\tau_k$, appearing in Eq. (\ref{Gaussianansatz}), is a wavevector dependent relaxation time.
According to Eq. (\ref{Gaussianansatz}), the real and imaginary parts of the Laplace transform of the memory function are given by, respectively \cite{Ailawadi,Hansen},
\begin{equation}
k^2\phi^{'}(k,\omega) = [\omega_L^2(k) - \langle\omega_k^2\rangle]\tau_ke^{-\tau_k^2\omega^2/\pi}
\label{realphi}
\end{equation}
and
\begin{equation}
k^2\phi^{''}(k,\omega) = \frac{2\tau_k}{\sqrt{\pi}}[\omega_L^2(k) - \langle\omega_k^2\rangle]D(\tau_k\omega/\sqrt{\pi})\,,
\label{imagphi}
\end{equation}
where the Dawson function $D(x) = \exp(-x^2)\int_0^x\exp(y^2)dy$
\cite{Dawson}.  

The quality of the Gaussian model has been previously identified for the Lennard-Jones fluid \cite{Ailawadi,Schepper} and by Hansen et al. in a pioneering study of the  One Component Plasma (OCP) \cite{Hansen}; it has also been applied to experimental data for weakly coupled plasma produced by arc jets \cite{Gregori2002}.
However, because of the difficulty of conducting highly accurate numerical simulations at the time of the previous investigations, a detailed, conclusive comparison of the model in Eq. (\ref{siiequation}) with the results of Molecular Dynamics (MD) simulations was not possible for those systems.
Here, with the aid of modern computing facilities, we have conducted accurate, large scale MD simulations for $S_{ii}(k,\omega)$ across a 
wide range of thermodynamic conditions.  We find that the Gaussian model matches the
MD data for the Yukawa system very well for all thermodynamic conditions we have examined in our simulations.

\subsection{Physical discussion of model for $S_{ii}(k,\omega)$}
\label{theory}
The structure factor in the hydrodynamic regime, as given in Eq. (\ref{siihydro}), can be derived from the longitudinal component of the linearized Navier Stokes equation,
\begin{equation}
\frac{d}{dt}J(\br,t) = -\frac{1}{m}\nabla P(\br,t) + \eta_l\nabla^2J(\br,t)\,,
\label{hydro1}
\end{equation}
where $J(\br,t)$ is the longitudinal current density and $P(\br,t)$ is the pressure.
Similarly, Eq. (\ref{siiequation}) can be derived from a generalized version of Eq. (\ref{hydro1}) 
(see \cite{Ailawadi} for more details),
\begin{align}
\frac{d}{dt}J(\br,t) = &-\frac{1}{m}\nabla \int d{\br}^{'} \frac{\delta P(\br,t)}{\delta n(\br^{'},t)}\delta n(\br^{'},t)\,.  \nonumber \\
&+\nabla^2 \int_0^t \int dsd{\br}^{'}\phi(\br - {\br}^{'},t-s)J({\br}^{'},s)\,,
\label{generalhydro}
\end{align}
where $n(\br,t)$ is the number density.
This generalization is motivated in the following way.  At small length scales, the validity of the conventional hydrodynamic description can be expected to break down.   Specifically, in the Navier Stokes description of Eq. (\ref{hydro1}), both the pressure term and viscosity term are local in space and time.
The generalization in Eq. (\ref{generalhydro}) includes the non-local behavior that is essential at small length scales in two ways.
Firstly, it is assumed that a change in pressure at a position $\br$ should not be determined completely by density fluctuations at the same position $\br$ but also by density fluctuations at neighbouring positions.  
This means that the pressure gradient due to a density gradient is non-local (hence the functional derivative appearing in Eq. (\ref{generalhydro})).
Secondly, the viscosity is made to be non-local in space and time to model the viscoelastic effects in a real liquid.  The memory function $\phi(\br,t)$ that models these viscoelastic effects describes the delayed response of the longitudinal part of the stress tensor to a change in the rate of shear \cite{Ailawadi}.  In Eq. (\ref{Gaussianansatz}), this response is modeled by a single relaxation time $\tau_k$.
The requirement that the model reproduces the result obtained from the Navier-Stokes equations in the hydrodynamic limit gives a relation between the long wavelength behavior of this relaxation time and the kinematic viscosity $\eta_l$ \cite{Ailawadi},
\begin{equation}
\eta_l = mn \lim_{k \to 0}[\omega_L^2(k) - \langle\omega_k^2\rangle]\tau_k/k^2 \,,
\label{hydrolimit}
\end{equation}
where $\eta_l = (\frac{4}{3}\eta + \zeta)/mn$, with $\eta$ and $\zeta$ the shear and bulk viscosities respectively. 

The generalization included in Eq. (\ref{generalhydro}) leads to the expression in Eq. (\ref{siiequation}) for the dynamical structure factor (see e.g. \cite{Ailawadi}).  All that remains is to specify the memory function.  As discussed in Sec. \ref{model}, here we choose a Gaussian memory function, as this is the simplest model that previous studies have suggested gives a good description of the dynamics of classical fluids. 
 We find that this choice yields a model of the dynamical structure factor that matches the MD data for the Yukawa system remarkably well.

\section{Results and Analysis}
\label{results}

The Gaussian memory function model given in Eqs. (\ref{siiequation}), (\ref{realphi}) and (\ref{imagphi}) requires values for $\langle\omega_k^2\rangle$, $\omega_L^2(k)$ and $\tau_k$ for each $k$. 
Since all three of these parameters are in general unknown, we have fitted them to the MD spectrum of $S_{ii}(k,\omega)$ using the least squares method.  
 That is to say, for each $k$ value for which we have computed $S_{ii}(k,\omega)$ with MD (these are the $k$ values compatible with the periodic boundary conditions in our simulations), we fit the model to the MD spectrum of $S_{ii}(k,\omega)$.
When this is done, the model reproduces the MD data very accurately for all $\Gamma$ and $\alpha$ values; in Sec. \ref{comparison} we show that this is the case for small, intermediate and large $k$ values (see also \cite{supplement}).

\begin{figure}[h!]
\includegraphics{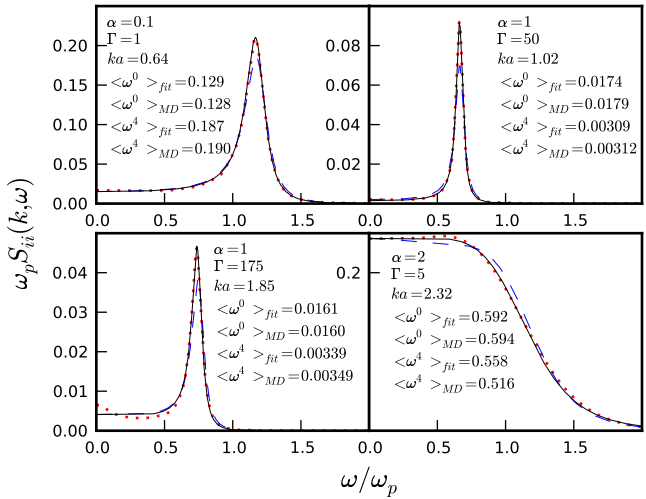}
\caption{(color online) Comparison between the Gaussian model when only the parameter $\tau_k$ is fitted to the MD spectrum (dashed line), and when all three parameters are fitted (solid line) for four separate cases.  The MD results are given by the dots.}
\label{fitplot}
\end{figure}

The three parameter fit is the correct way to compare the Gaussian memory function model to the MD spectrum of $S_{ii}(k,\omega)$.  This is true despite the fact that two of the parameters, $\langle\omega_k^2\rangle$ and $\omega_L^2(k)$, can in principle be obtained by computing $S_{ii}(k)$ (or equivalently the radial distribution function $g(r)$ \cite{HansenMcdonald}) with MD and using the formulae given in the Appendix \ref{appendix2}.  When obtained from MD in this way, these two parameters are subject to numerical incertainty.  Therefore, one would expect that constraining $\langle\omega_k^2\rangle$ and $\omega_L^2(k)$ - and therefore fitting the model to the MD spectrum using only a single parameter $\tau_k$ \cite{Hansen,Ailawadi,Gregori2002} - would result in poorer fits and larger errors.  In Fig \ref{fitplot}, we show that in general this is indeed the case.

\begin{figure}[h!]
\includegraphics{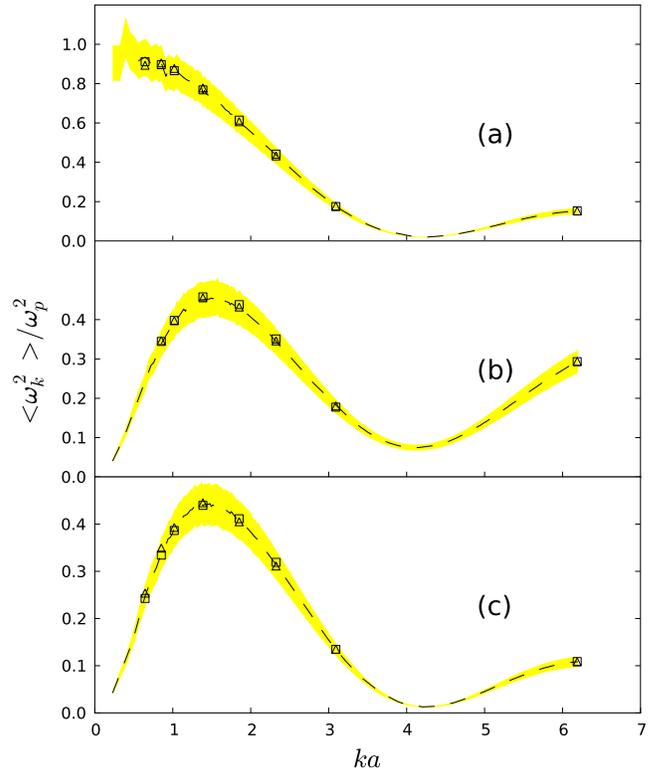}
\caption{(color online) Comparison between $\langle\omega_k^2\rangle$ as computed from MD using the formulae in the Appendix (dashed line, with $10\%$ error band), and the values obtained from the three parameter fit of the Gaussian memory function model (triangles) and the viscoelastic model (squares) for three different plasma conditions. (a) $\Gamma = 120$, $\alpha = 0.1$, (b) $\Gamma = 50$, $\alpha = 1$, (c) $\Gamma = 175$, $\alpha = 1$.}
\label{wkfig}
\end{figure}

\begin{figure}[h!]
\includegraphics{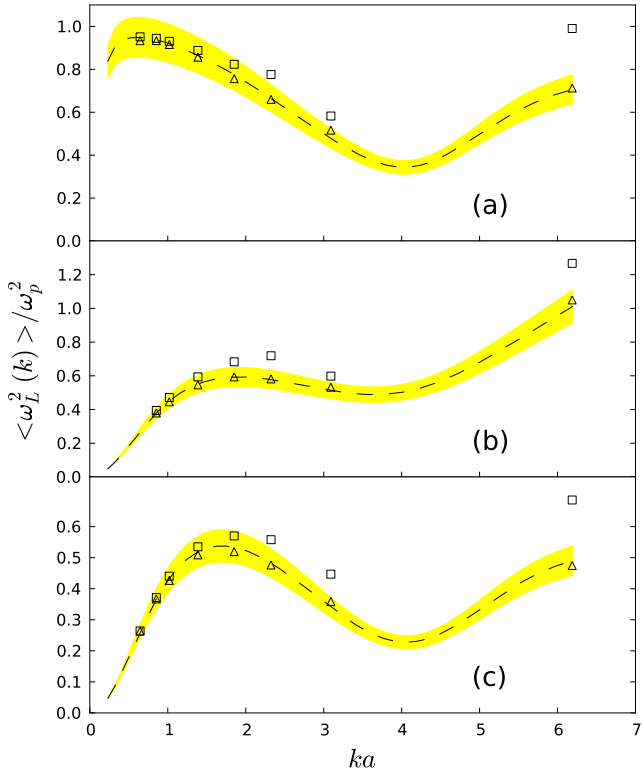}
\caption{(color online) Comparison between $\langle\omega_L^2(k)\rangle$ as computed from MD using the formulae in the Appendix (dashed line, with $10\%$ error band), and the values obtained from the three parameter fit of the Gaussian memory function model (triangles) and the viscoelastic model (squares) for three different plasma conditions. (a) $\Gamma = 120$, $\alpha = 0.1$, (b) $\Gamma = 50$, $\alpha = 1$, (c) $\Gamma = 175$, $\alpha = 1$.}
\label{wlfig}
\end{figure}

The validity of the three parameter fit can be confirmed by comparing the fitted values of the two parameters $\langle\omega_k^2\rangle$ and $\omega_L^2(k)$ to their values when instead computed with MD as described above.  As shown in  Figs. \ref{wkfig} and \ref{wlfig}, the parameters $\langle\omega_k^2\rangle$ and $\omega_L^2(k)$ obtained from the fit to the MD spectrum of $S_{ii}(k,\omega)$ agree very well (within $10\%$) with those computed from the MD $g(r)$ and $S_{ii}(k)$.
This is only the case because the model works so well.  For example, as shown in Fig. \ref{wlfig}, if an exponential rather than Gaussian memory function is used (this is known as the viscoelastic model and is discussed in Sec. \ref{viscoelastic}), the numerical values obtained for $\omega_L^2(k)$ by fitting the model with three parameters do not agree well with those computed from the MD $g(r)$ and $S_{ii}(k)$.
In the remainder of the paper, we present only the results for the Gaussian memory function model with three fitting parameters; the one parameter fits are irrelevant as their comparison with the MD data for $S_{ii}(k,\omega)$ is not indicative of the quality of the model.

\subsection{Comparison between model and MD simulations}
\label{comparison}

\begin{figure}[h!]
\includegraphics{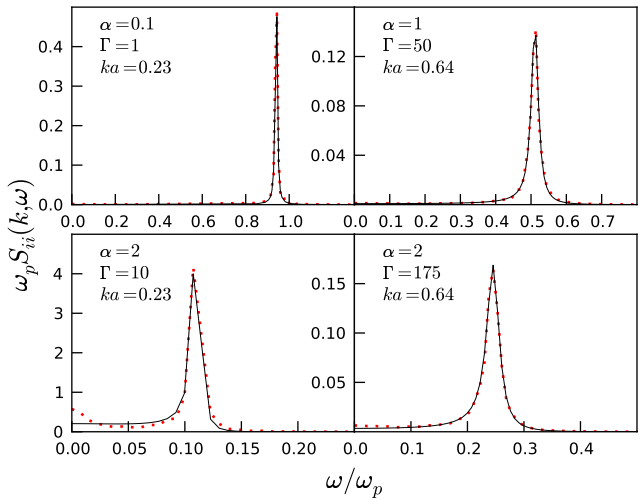}
\caption{(color online) Comparison between the MD data for $S_{ii}(k,\omega)$ (dots) and the Gaussian memory function model with three fitting parameters (solid line) for small $ka$ values.}
\label{smallk}
\end{figure}

\begin{figure}[h!]
\includegraphics{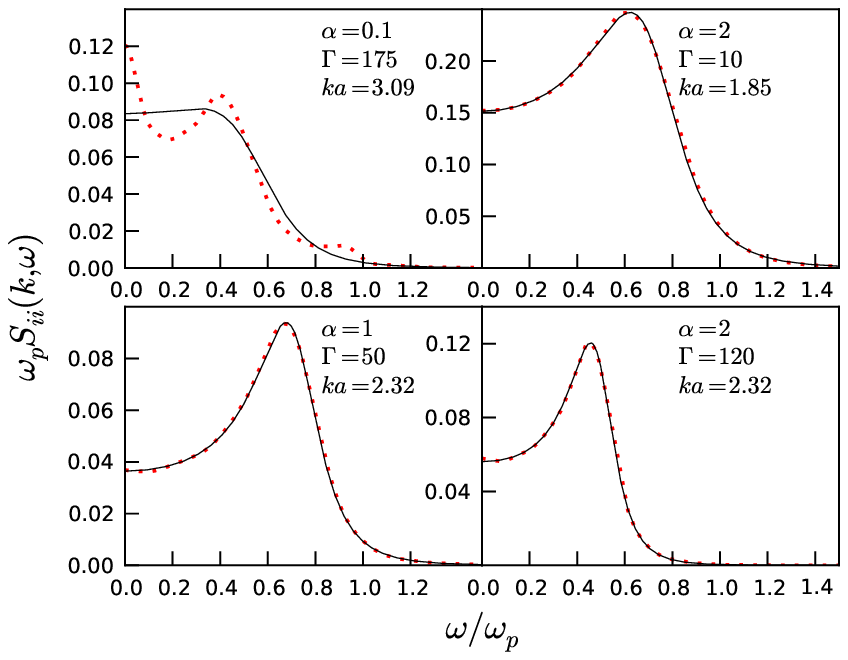}
\caption{(color online) Comparison between the MD data for $S_{ii}(k,\omega)$ (dots) and the Gaussian memory function model with three fitting parameters (solid line) for intermediate $ka$ values.}
\label{midk}
\end{figure}

\begin{figure}[h!]
\includegraphics{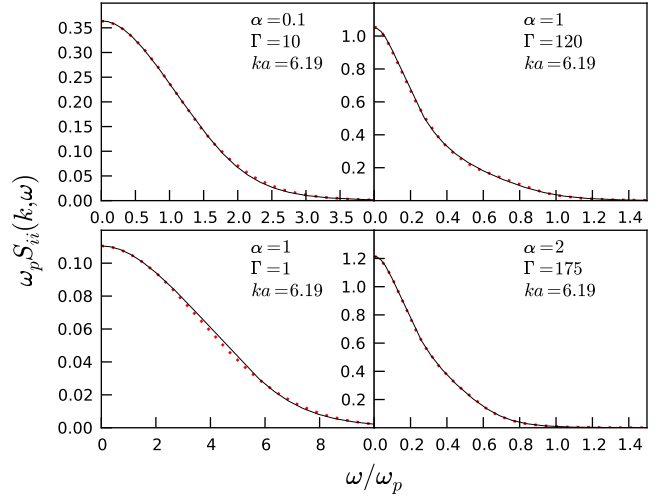}
\caption{(color online) Comparison between the MD data for $S_{ii}(k,\omega)$ (dots) and the Gaussian memory function model with three fitting parameters (solid line) for large $ka$ values.}
\label{largek}
\end{figure}

We find that in general the Gaussian memory function model reproduces the MD data very well for all of the $\Gamma$  ($1$,$5$,$10$,$50$,$120$,$175$) and $\alpha$ ($0.1$,$1$ and $2$) values we have considered, at all $k$ values (our simulations are for $ka = 0.23 - 6.19$).  Extended figures of our complete MD results are available as supplementary material \cite{supplement}; here, in Figs. \ref{smallk} - \ref{largek}, we show only a selection of these complete results at small, intermediate, and large $k$ respectively.

At small $k$ values (Fig. \ref{smallk}), for all $\alpha$ and $\Gamma$, the MD data shows a clear ion-acoustic (or Brillouin) peak that represents a damped sound wave in the plasma.
In this regime, the model extends the conventional hydrodynamic description to finite $k$ values. Specifically, 
the generalised sound speed along with the imaginary part of $\phi(k,\omega)$ correct for the fact that the position of the peak does not vary linearly with $k$ as in the
hydrodynamic description \cite{Mithen}, and the real part of $\phi(k,\omega)$ corrects for the width.  

At intermediate $k$ values (Fig. \ref{midk}), the model gives a surprisingly accurate account of both the width and position of the ion acoustic peak.  
This is particularly true for $\Gamma \leq 50$.  
For higher $\Gamma$ values, the MD data does in some cases show additional structure which the model cannot recreate.  
In particular, for $\alpha = 0.1 \mbox{ and } 1$, a two peak structure
is visible for $ka = 2.32$ and a three peak structure for $ka = 3.09$  (e.g. Fig. \ref{midk}, top left).  
The small peak just below $\omega_p$ for $ka = 3.09$ is of particular interest - it does not appear to have been seen or commented upon in previous MD calculations.  
We note that this peak is distinct from the higher harmonic peaks reported in \cite{Hartmann}.  In fact, at $\kappa = 0.1$ only, we do see signs of a second harmonic peak, at a frequency close to $2\omega_p$.  We have neglected this harmonic peak in our analysis, since we find it to be more than than 3 orders of magnitude smaller than the main features in the spectrum of $S_{ii}(k,\omega)$, in good agreement with \cite{Hartmann}.  On the other hand, the peak shown in Fig. \ref{midk} (top left) is of the same order of magnitude as the main features of $S_{ii}(k,\omega)$.
We believe that this peak is due to microscopic `caging' effects (e.g. \cite{HansenMcdonald,BalucaniZoppi}).  That is, at these lengthscales, the relatively high frequency oscillations of individual particles in the cages produced by their neighbors are imprinted on $S_{ii}(k,\omega)$.
We note that although the model does not fully capture the additional structure in 
the MD data for these conditions, on average it does give a good account of the overall shape of the spectrum.

\begin{figure}[h!]
\includegraphics{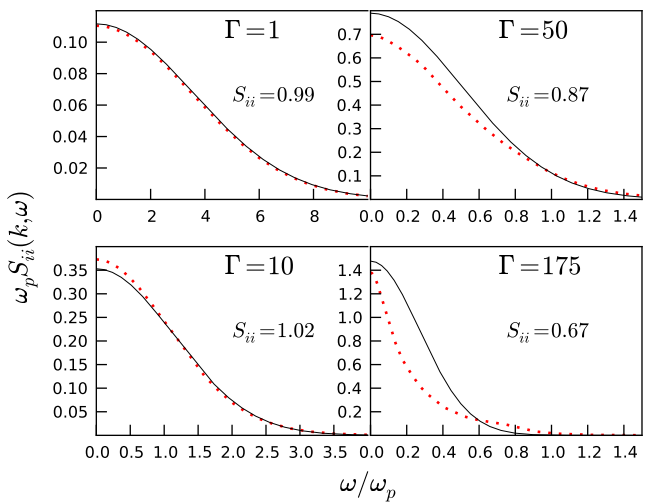}
\caption{(color online) Comparison between the MD data for $S_{ii}(k,\omega)$ for $\alpha = 1$ and $ka = 6.19$ (dots) and the ideal gas limit given by Eq. (\ref{idealgas}) (solid line).  Also shown on each panel is the value of the static structure factor $S_{ii}(k)$ at $ka = 6.19$.}
\label{idealfig}
\end{figure}

At large $k$ values (Fig. \ref{largek}), $S_{ii}(k,\omega)$ reduces to a single peak at $\omega = 0$.
In this regime, the model reproduces the MD data very accurately in all cases.
As $k$ increases, $S_{ii}(k,\omega)$ should tend to its ideal gas limit $S_{ii}^{0}(k,\omega)$, which is independent of $\alpha$ \cite{Hansen,HansenMcdonald},
\begin{equation}
S_{ii}^{0}(k,\omega) = \left(\frac{m}{2\pi k_B T k^2}\right)^{1/2}\exp\left(-\frac{m\omega^2}{2k_BTk^2}\right)\,.
\label{idealgas}
\end{equation}
As shown in Fig. \ref{idealfig}, at constant $\alpha$,  as $\Gamma$ increases $S_{ii}(k,\omega)$ converges 
more slowly towards $S_{ii}^{0}(k,\omega)$.
Indeed, at the highest $k$ value we have considered in our MD simulations ($ka = 6.19$), 
the MD result only compares well to its ideal gas limit for $\Gamma \leq 10$ (see Fig. \ref{idealfig}).  
We note that the discrepancy between $S_{ii}(k,\omega)$ and
its ideal gas limit can more readily be seen by looking at the MD data for the static structure factor $S_{ii}(k)$
; the ideal gas limit will only be approximated at $k$ values for which $S_{ii}(k) \approx 1$ (since $S_{ii}^{0}(k) = 1$).

In any case, as shown in Fig. \ref{largek}, the Gaussian model compares very well to the MD data at our highest $k$ value of $ka = 6.19$, regardless of whether or not this $k$ value is sufficiently large for $S_{ii}(k,\omega)$ to be close to its ideal gas limit.

\subsection{Hydrodynamic limit}
\label{hydrolimitsection}
In previous investigations (e.g. \cite{Ailawadi}), Eq. (\ref{hydrolimit}) was used to infer the kinematic viscosity from the long wavelength behavior of the relaxation time $\tau_k$ appearing in the memory function.
For the Yukawa system, in principle this could be used to determine the shear viscosity (the bulk viscosity is in general negligible in comparison with the shear viscosity for the Yukawa system \cite{Salin}).
However, due to the inaccuracy inherent in measuring the width of the (very narrow) ion acoustic peak obtained from 
the MD simulations at small $k$ values, we find that this method is of little practical use compared to other approaches to determining the viscosity.  These alternative approaches include utilizing the Green-Kubo relation for the shear stress autocorrelation function \cite{Saigo}, non-equilibrium molecular dynamics methods \cite{Donko2000}, and computation of the transverse current autocorrelation function \cite{Donko2010}.

Along with the generalized viscosity, as discussed in Sec. \ref{model}, in the hydrodynamic limit $k \rightarrow 0$ the generalized sound speed $c_s(k) = \sqrt{\langle\omega_k^2\rangle/k^2}$ reduces to the conventional (isothermal) sound speed $c_s$.
The small $k$ behaviour of the generalized viscosity and sound speed thus ensure that using the Gaussian ansatz for the memory function in Eq. (\ref{siiequation}) gives a result that is 
compatible with the result obtained from the linearised Navier Stokes equations \cite{HansenMcdonald} {\it when thermal fluctuations are neglected}.
To be clear, Eq. (\ref{siiequation}) is an entirely general (i.e. exact) representation of $S_{ii}(k,\omega)$. The effective neglect of thermal fluctuations is made by assuming the ansatz in Eq. (\ref{Gaussianansatz}). That is to say, in the case of the Gaussian ansatz it is instructive to think of the memory function as a sort of generalized viscosity.  There is no term in the memory function that represents the effects of temperature fluctuations i.e. a generalized (or indeed non-generalized) thermal conductivity. 

It is straightforward to modify Eq. (\ref{Gaussianansatz}) so that the result from the Navier Stokes equations including temperature fluctuations is recovered in the hydrodynamic limit (see e.g. \cite{BoonYip,Ailawadi}).
The simplest extension involves maintaining a generalized sound speed and viscosity, and adding the (non-generalized) thermal conductivity contribution obtained from conventional hydrodynamics (the Navier-Stokes equations) as an additional term in the memory function.  In a more involved scheme, this additional contribution can also be generalized \cite{Schepper,BoonYip}.

For the Yukawa system with the $\Gamma$ and $\alpha$ values we have considered here, including in the memory function the effects of thermal fluctuations is unnecessary.
This is because the ratio of specific heats, $\gamma$, is very close to $1$, as indicated by the absence of a Rayleigh peak at $\omega = 0$ for small $k$ in the MD data (Fig. \ref{smallk}), as well as previous equation of state calculations \cite{Hamaguchi}.  The only cases in which this peak - which represents a diffusive thermal mode - is not negligible is for the more weakly coupled ($\Gamma \leq 10$) systems at $\alpha = 2$ (see Fig. \ref{smallk}, bottom left).  As expected, the model does not capture this peak in the MD data.

The fact that $\gamma \approx 1$ for the Yukawa system with the $\Gamma$ and $\alpha$ values considered here is certainly a reason why the Gaussian memory function works so well.  Indeed, the ansatz in Eq. (\ref{Gaussianansatz}) would not be expected to work as well when the ratio of specific heats $\gamma$ is noticeably different from unity \cite{BalucaniZoppi}; this includes the Yukawa system for $\Gamma \ll 1$.

\subsection{Comparison with viscoelastic model}
\label{viscoelastic}
Given the excellent agreement between the MD data and the Gaussian memory function model, we have not found it necessary to undertake an exhaustive comparison with the numerous other forms of memory function proposed in the literature \cite{BoonYip}.  
However, here we briefly comment on another widely studied and used ansatz for the memory function
\begin{align}
k^2\phi(k,t) &= k^2\phi(k,0)\exp(-t/\tau^V_k) \nonumber \\
             &= [\omega_L^2(k) - \langle\omega_k^2\rangle]\exp(-t/\tau^V_k)\,.
\label{viscoansatz}
\end{align}
When combined with Eq. (\ref{siiequation}), Eq. (\ref{viscoansatz}) - which represents the simplest assumption that can be made about the time dependence of the memory function - is known as the viscoelastic model \cite{BalucaniZoppi}.

\begin{figure}[h!]
\includegraphics{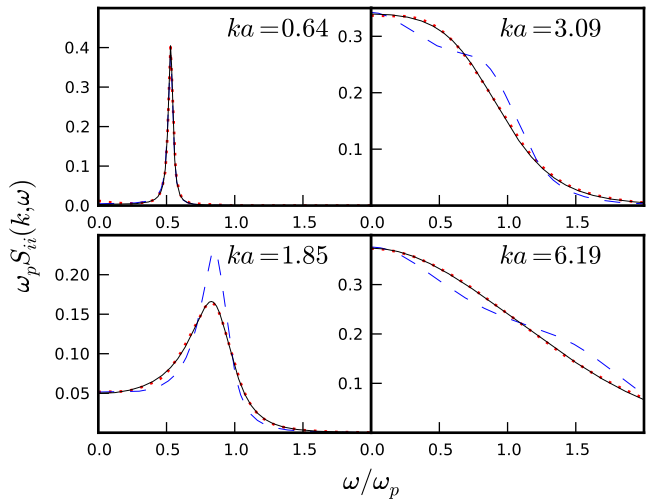}
\caption{(color online) A sample of our MD results for $S_{ii}(k,\omega)$ at 
$\Gamma = 10$, $\alpha = 1$ (dots) contrasting the results of the model in
Eq. (\ref{siiequation}) for exponential (dashed line) and
Gaussian (solid line) memory functions.}
\label{fig1}
\end{figure}

As indicated in Fig. \ref{fig1} and discussed in detail elsewhere \cite{BalucaniZoppi,Schepper,Ailawadi}, 
the viscoelastic model cannot capture the shape of $S_{ii}(k,\omega)$ across a large range of $k$ values.  
While the model works well at small $k$ (indeed, for the viscoelastic model the results of isothermal hydrodynamics are again recovered, with a relation between the relaxation time $\tau^V_k$ and the kinematic viscosity similar to Eq. (\ref{hydrolimit})), the model tends to predict rather more structure in $S_{ii}(k,\omega)$ than is evident in the MD data (Fig. \ref{fig1}).  Clearly then the Gaussian memory function is vastly superior to the exponential one.

\subsection{Discussion of the relaxation time $\tau_k$}

\begin{figure}[h!]
\includegraphics{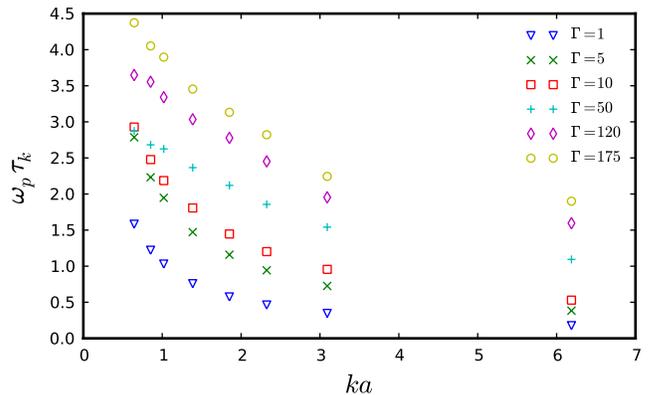}
\caption{(color online) The relaxation time $\tau_k$ as determined from the fit of the Gaussian model to the MD spectrum of $S_{ii}(k,\omega)$ for $\alpha = 2$ and a range of $\Gamma$ values.}
\label{taukplot}
\end{figure}

 Fig. \ref{taukplot} shows the relaxation time $\tau_k$ as determined from the fit of the Gaussian model to the MD spectrum of $S_{ii}(k,\omega)$ for $\alpha = 2$.  As shown in Fig. \ref{taukplot}, we find that as $k$ increases, $\tau_k$ decreases.  This agrees qualitatively with e.g. the behavior of the relaxation time determined for the Lennard-Jones fluid in previous investigations \cite{Ailawadi,Levesque}.  
One certainly expects that at decreasing wavevectors, the relaxation time should increase: as $k \rightarrow 0$, the memory function should decay fast enough to guarantee the validity of the Markovian approximation, which itself is related to the fulfillment of the conservation laws  \cite{BalucaniZoppi}.

In our investigation, we find that at the very smallest $k$ values accessible to our simulations (i.e. below $ka = 0.64$, which is the minimum $k$ value shown in Fig. \ref{taukplot}), the numerical value of $\tau_k$ is difficult to extract from the MD spectrum reliably, and therefore it is not possible to examine the exact $k \rightarrow 0$ behavior of the relaxation time.  That is to say, the fitted value of $\tau_k$ at these small $k$ values does not connect smoothly to the values at higher $k$ values; this is because the spectrum $S_{ii}(k,\omega)$ consists of a very sharp peak, for which it is difficult to accurately determine the parameters in the Gaussian model (see also Sec. \ref{hydrolimitsection}).

Physically, the relaxation time $\tau_k$ controls the specific collective behavior of the system: for times $t \ll \tau_k$ the system responds `elastically' (i.e. like a `frozen' solid-like system), wheras for times $t \gg \tau_k$ the viscous mechanisms set in and reveal the inherent dynamic disorder \cite{BalucaniZoppi}.  Therefore, the decrease in $\tau_k$ as $k$ increases corresponds physically to the fact that at increasingly short lengthscales, viscous behavior is observed at increasingly short timescales.

\subsection{Applicability to x-ray scattering experiments}
\label{applicability}
In a previous work \cite{Mithen}, it was shown that the conventional hydrodynamic description (i.e. Eq. (\ref{siihydro})) is valid providing $k < k_{max}$ , where $k_{max}\lambda_{s} \simeq 0.43$.  This means that experiments designed to measure $S_{ii}(k,\omega)$ \cite{Glenzer} at $k$ values below $k_{max}$ can in principle be used to determine transport (e.g. viscosity) and thermodynamic properties (e.g. compressibility) of dense plasmas.

At $k$ values larger than $k_{max}$, our results show that the Gaussian memory function model extends the conventional hydrodynamic description very satisfactorily.
Thus experiments for $k > k_{max}$ measure the generalized quantities appearing in the memory function model of Eq. (\ref{siiequation}). 

Present x-ray scattering experiments are also concerned with diagnosing the density and temperature of dense plasmas \cite{Glenzer}.  
For this task theoretical models for how $S_{ii}(k,\omega)$ depends on density and temperature are required.
 In the Yukawa system, the density and temperature are encoded in $\Gamma$ and $\alpha$.  Thus here we briefly look qualitatively at how $S_{ii}(k,\omega)$ changes with $\Gamma$ and $\alpha$: this gives an indication of how the experimental scattering cross section should vary with density and temperature.
We restrict ourselves to the region of $k$ values for which $S_{ii}(k,\omega)$ shows a clear ion-acoustic peak, since then its description reduces to the position, width and height of this peak.

\begin{figure}[h!]
\includegraphics{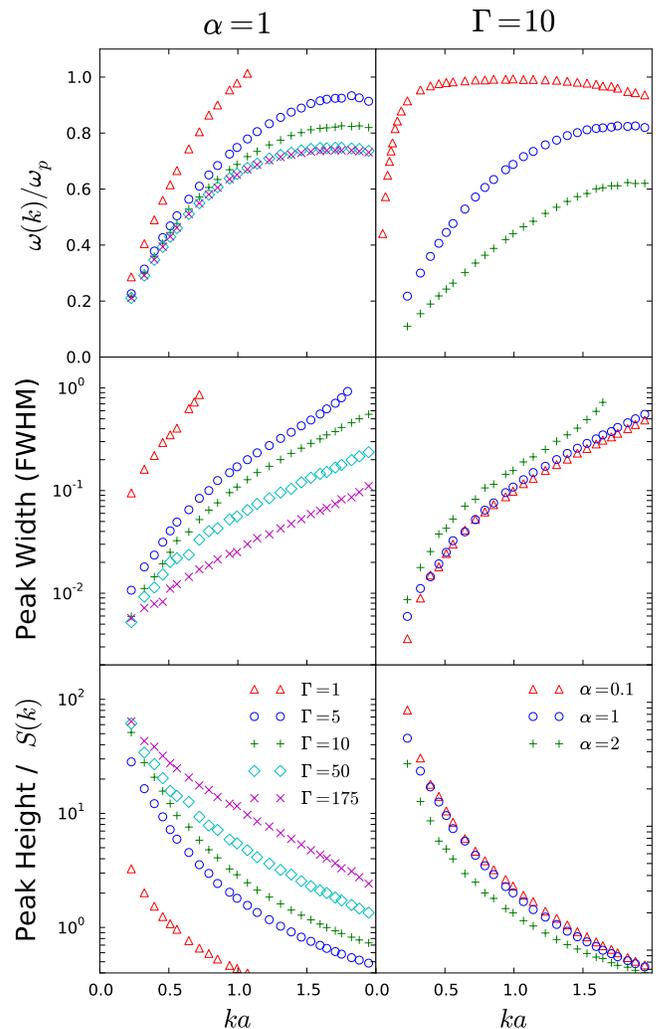}
\caption{(color online) Variation of position $\omega(k)$, width (log scale) and height (log scale) of ion-acoustic peak with reduced wavenumber $ka$.  The left panels are for $\alpha = 1$ and a range of $\Gamma$ values, and the right panels are for $\Gamma = 10$ and a range of $\alpha$ values.}
\label{bigplot}
\end{figure}

Fig. \ref{bigplot} shows how the position, width and height of the ion-acoustic peak as extracted from our MD simulations vary with reduced wavenumber $ka$ for a number of $\Gamma$ and $\alpha$ values.  As shown in the top left panel of Fig. \ref{bigplot}, the dependence on $ka$ of the ion-acoustic peak position is almost identical for a large range of $\Gamma$ values (i.e. $\Gamma = 50 - 175$). The peak width and height do show more discernible differences for these $\Gamma$ values.  
At smaller $\Gamma$ values ($\Gamma = 1$,$5$ and $10$), the differences in the position, width and height of the peak are greater.

At constant $\Gamma$ (right panels of Fig. \ref{bigplot}), the peak position is rather different for $\alpha = 0.1$,$1$ and $2$. In this case, the width and height are more similar, particularly for $\alpha = 0.1$ and $\alpha = 1$.

We expect that a given experiment will be able to determine peak position, width and height at a specific wavenumber (determined by the scattering angle and x-ray wavelength \cite{Glenzer}). The extraction of $\Gamma$ and $\alpha$ values could then be done by using these experimental results in conjunction with a set of three plots as shown in Fig. \ref{bigplot}.

Of course, our discussion in this section assumes that  a real physical plasma at a certain density, temperature and (average) ionization state can be described by the Yukawa system.  While in principle this mapping could be attempted for any given values of these plasma parameters, our main interest at present concerns the dense (approximately solid density), liquid-like plasmas at temperatures of $\approx 10eV$ that can be created in high power laser experiments \cite{Glenzer}.  Recently, a method for mapping the physical parameters of these states to the Yukawa model (i.e. determination of $\Gamma$ and $\alpha$) has been suggested \cite{Murillo}.  Therefore, we expect that the results we have obtained for the Yukawa system are certainly relevant for future experiments that will measure ion dynamics of these extreme states of matter.

\section{Concluding Comments}
\label{conclusion}
The Gaussian memory function model is an extremely good representation of the dynamical structure factor $S_{ii}(k,\omega)$ of the Yukawa system for a wide range of thermodynamic conditions.  
The model very accurately reproduces the spectrum of $S_{ii}(k,\omega)$ from MD in terms of just 3 parameters and, as such, it is a useful way of accurately condensing or representing such data.
This conclusion was only possible because of the highly accurate MD data presented in this paper.  The model can be used by fitting either a single parameter or three parameters to the spectrum of $S_{ii}(k,\omega)$ at a particular $k$ value; in the latter case, the small numerical inaccuracies that arise in the MD simulations can be accounted for.

Why exactly this form of memory function should work so well is an interesting question that certainly merits further investigation.  
Other memory function models, such as the viscoelastic model (an exponential memory function) do not compare well to the MD data for a wide range of $k$ values.  It is possible that the reason a faster decaying (compared to exponential) Gaussian works well is related to the chaotic nature of classical systems - this is reflected in the relatively short `memory' of the system.

Since the Yukawa system can describe ion-ion interactions in a plasma, our results are applicable to future x-ray scattering experiments that will attempt to measure ion dynamics in dense plasmas \cite{Gregori}. In particular, our MD results for the position, width and height of the ion-acoustic peak could be used to infer the thermodynamic conditions of dense plasmas.  

\section{Acknowledgements}

This work was supported by the John Fell Fund at the University of Oxford and by EPSRC
grant no. EP/G007187/1.  The work of J.D. was performed for the U.S. Department of Energy
by Los Alamos National Laboratory under Contract No.
DE-AC52-06NA25396. J.D. and J.P.M. gratefully acknowledge the support of the US Department of Energy through the LANL/LDRD Program for this work.

\appendix*

\section{Frequency moments of $S_{ii}(k,\omega)$}
\label{appendix2}
The wavevector dependent quantities,
\begin{equation}
\langle\omega_k^2\rangle = \frac{\langle\omega^2\rangle}{\langle\omega^0\rangle}\,,
\end{equation}
and
\begin{equation}
\omega_L^2(k) = \frac{\langle\omega^4\rangle}{\langle\omega^2\rangle}\,,
\end{equation}
are given in terms of
the frequency moments of $S_{ii}(k,\omega)$, defined as
\begin{equation}
\langle\omega^n\rangle = \int_{-\infty}^{\infty}\omega^nS_{ii}(k,\omega)d\omega\,.
\end{equation}
The zeroth moment of $S_{ii}(k,\omega)$ gives the static structure factor $S_{ii}(k)$
\begin{equation}
\langle\omega^{0}\rangle = S_{ii}(k)\,.
\end{equation}
The second moment is
\begin{equation}
\frac{\langle\omega^{2}\rangle}{\omega_p^2} = \frac{q^2}{3\Gamma}\,,
\end{equation}
where $q = ka$ is the reduced wavevector ($a = (3/(4\pi n))^{1/3}$ is the Wigner-Seitz radius) and $\omega_p = \sqrt{(Z^2e^2n)/(\epsilon_0m)}$ is the (ion) plasma frequency.  
The fourth moment is (see \cite{BalucaniZoppi}, Eq. (1.137))
\begin{equation}
\frac{\langle\omega^4\rangle}{\omega_p^4} = \frac{1}{3\Gamma}\left[\frac{q^4}{\Gamma} + q^2\Omega_E^2 - q^2M(q\bar{r},\alpha\bar{r})\right]\,.
\label{4thmoment}
\end{equation}
Here $\bar{r} = r / a$, the Einstein frequency $\Omega_E$ is given by
\begin{equation}
\Omega_E^2 = \frac{\alpha^{2}}{3}\int_{0}^{\infty}\bar{r}\exp(-\alpha\bar{r})g(\bar{r})d\bar{r}\,,
\end{equation}
and
\begin{align}
M(x,y) &= \int_0^{\infty}\frac{1}{\bar{r}}g(\bar{r})\exp(-y)\left[2\left(\frac{y^2}{3} + y + 1\right)\times\right.\nonumber\\
&\left.\left(\frac{\sin x}{x} + \frac{3\cos x}{x^2} - \frac{3\sin x}{x^3}\right) + \frac{y^2\sin x}{3x}\right]d\bar{r}\,.
\label{4thmoment2}
\end{align}
Eqs. \ref{4thmoment} - \ref{4thmoment2} give an exact expression for the fourth moment for the Yukawa one component plasma.

\end{document}